\documentclass[aps,prd,reprint,nofootinbib,twocolumn,superscriptaddress,preprintnumbers]{revtex4-2}

\usepackage{amsthm}
\usepackage{amsmath}
\usepackage{graphicx}
\usepackage{slashed}
\usepackage{amssymb}
\usepackage[colorlinks=True, citecolor=blue, urlcolor=blue, linkcolor=blue]{hyperref}
\usepackage{multirow}
\usepackage{orcidlink}
\usepackage{qcircuit}
\usepackage{physics}
\usepackage{wasysym}

\newcommand{\nn}{\nonumber}

\newcommand{\bea}{\begin{align}}
\newcommand{\eea}{\end{align}}

\begin{document}

\title{Quantum Error Correction Codes for Truncated SU(2) Lattice Gauge Theories}

\author{Xiaojun Yao}
\email{xjyao@uw.edu}
\affiliation{InQubator for Quantum Simulation, Department of Physics, University of Washington, Seattle, Washington 98195, USA}

%\date{\today}
\preprint{IQuS@UW-21-114}
\begin{abstract}
We construct two quantum error correction codes for pure SU(2) lattice gauge theory in the electric basis truncated at the electric flux $j_{\rm max}=1/2$, which are applicable on quasi-1D plaquette chains, 2D honeycomb and 3D triamond and hyperhoneycomb lattices. The first code converts Gauss's law at each vertex into a stabilizer while the second only uses half of the vertices and is locally the carbon code. Both codes are able to correct single-qubit errors.
The electric and magnetic terms in the SU(2) Hamiltonian are expressed in terms of logical gates in both codes. The logical-gate Hamiltonian in the first code exactly matches the spin Hamiltonian for gauge singlet states found in previous work.
\end{abstract}

\maketitle

\section{Introduction}
Gauge theory lays the theoretical foundation for the Standard Model of particle physics and also describes many interesting models in condensed matter physics such as the toric code. Analytically solving generic gauge theories, in particular non-Abelian gauge theories such as the Quantum Chromodynamics, is very challenging. Most numerical studies on the lattice are limited to static observables due to the real-time sign problem in the path integral and the exponential growth of the Hilbert space with the lattice size. 

A quantum computer can in principle cure both problems. With the rapid development of quantum technology, quantum simulation of lattice gauge theory has attracted widespread interest in the physics community~\cite{Farrell:2022wyt,Farrell:2022vyh,Osborne:2022jxq,Farrell:2023fgd,Dempsey:2023fvm,Farrell:2024fit,Illa:2024kmf,Farrell:2024mgu,Mueller:2024mmk,Lee:2024jnt,Fontana:2024rux,Schuhmacher:2025ehh,Yang:2025edn,Turro:2025sec,Balaji:2025afl,Balaji:2025yua,Ciavarella:2025bsg,Davoudi:2025rdv,Halimeh:2025ivn,Briceno:2025ptn,Xu:2025abo,Joshi:2025rha,Joshi:2025pgv,Miranda-Riaza:2025fus,Ciavarella:2025tdl,Kane:2025ybw,Tian:2025mbv,Cataldi:2025cyo,Santra:2025dsm}. In the Kogut-Susskind Hamiltonian formulation of lattice gauge theory, local Gauss's law constraints are imposed at every vertex, which physically means that the vertex state behaves as a singlet under gauge transformations. In practice, different treatments of Gauss's law have been used such as constructing singlet states at each vertex~\cite{Klco:2019evd,Ciavarella:2021nmj,Anishetty:2009nh,Raychowdhury:2019iki,Kadam:2022ipf,Kadam:2024ifg}, completely fixing the gauge~\cite{Mariani:2024osg,Grabowska:2024emw,Burbano:2024uvn}, using a different gauge than the temporal gauge~\cite{Li:2024ide,Yao:2025uxz}, and suppressing Gauss's law violations by adding penalty terms to the original Hamiltonian~\cite{Zohar:2012ay,Zohar:2012ts,Halimeh:2021vzf,Halimeh:2021lnv}. The spirit of these approaches is to address the challenge raised by Gauss's law and maintain gauge invariance of the theory. 

On a different perspective, Gauss's law can be viewed as an advantage of gauge theory for quantum simulation. Gauge invariance means redundancy in the description of the physical Hilbert space. In the Kogut-Susskind formulation, physical states are singlets under gauge transformations and the total Hilbert space contains both physical and unphysical states. Gauss's law picks up the gauge singlet sector. In this spirit, gauge theory has a natural redundancy for quantum error correction and Gauss's law can be used as a check for error syndromes. This line of thinking has been pursued by several groups, to detect errors that break Gauss's law as an oracle~\cite{Stryker:2018efp}, to construct error correction codes for fermionic quantum simulation~\cite{Chen:2022dox} and for lattice gauge theories with Abelian groups [$\mathbb{Z}_2$ and truncated U(1)]~\cite{Rajput:2021trn,Spagnoli:2024mib} and discrete groups~\cite{Carena:2024dzu}, to build approximate codes for continuous symmetries~\cite{Faist:2019ahr,Woods:2019fpy}, and to understand the relation in terms of quantum reference frames~\cite{Carrozza:2024smc}. Utilizing Gauss's law opens new ways of constructing error correction codes, which may be more natural for quantum simulations of lattice gauge theories than other generic codes~\cite{Shor:1995hbe,Laflamme:1996iw,Calderbank:1995dw,Steane:1996ghp,Kitaev:1997wr,Bravyi:1998sy,Bombin:2006sj,Pastawski:2015qua,Bravyi:2023qpn}, some of which are proposed and tested for general-purpose fault-tolerant quantum computing~\cite{Fowler:2012hwn,GoogleQuantumAIandCollaborators:2024efv,Paetznick:2024ztu,Yoder:2025ooz}. 

Here in this work, we take the first attempt to construct quantum error correction codes for non-Abelian lattice gauge theories by utilizing Gauss's law. In particular, we take pure SU(2) gauge theory as an example on lattices with only vertices to which at most three links are attached, including quasi-one-dimensional (quasi-1D) plaquette chains, 2D honeycomb~\cite{Muller:2023nnk} and 3D triamond~\cite{Kavaki:2024ijd} and hyperhoneycomb lattices~\cite{Illa:2025dou}. We use the electric basis and truncate the electric flux at $j_{\rm max}=1/2$. Two quantum error correction codes are constructed based on Gauss's law. The first code converts Gauss's law constraints at all vertices into stabilizers while the second only converts half.

The paper is organized as follows: We will briefly review the Kogut-Susskind Hamiltonian of pure SU(2) lattice gauge theory in Sec.~\ref{sec:lgt}. The first error correction code will be constructed in Sec.~\ref{sec:code1} for plaquette chains with aperiodic boundary conditions and extended to 2D and 3D in Sec.~\ref{sec:2D3D}. Then we will discuss the case of periodic boundary conditions in Sec.~\ref{sec:pbc}, where we will see the appearance of topological flux configurations. The construction of the second code will be given in Sec.~\ref{sec:code2}. Finally, we will summarize and draw conclusions in Sec.~\ref{sec:conclusions}.

\section{SU(2) Lattice Gauge Theory on Plaquette Chain}
\label{sec:lgt}
\begin{figure}[t]
\centering
\includegraphics[width=0.48\textwidth]{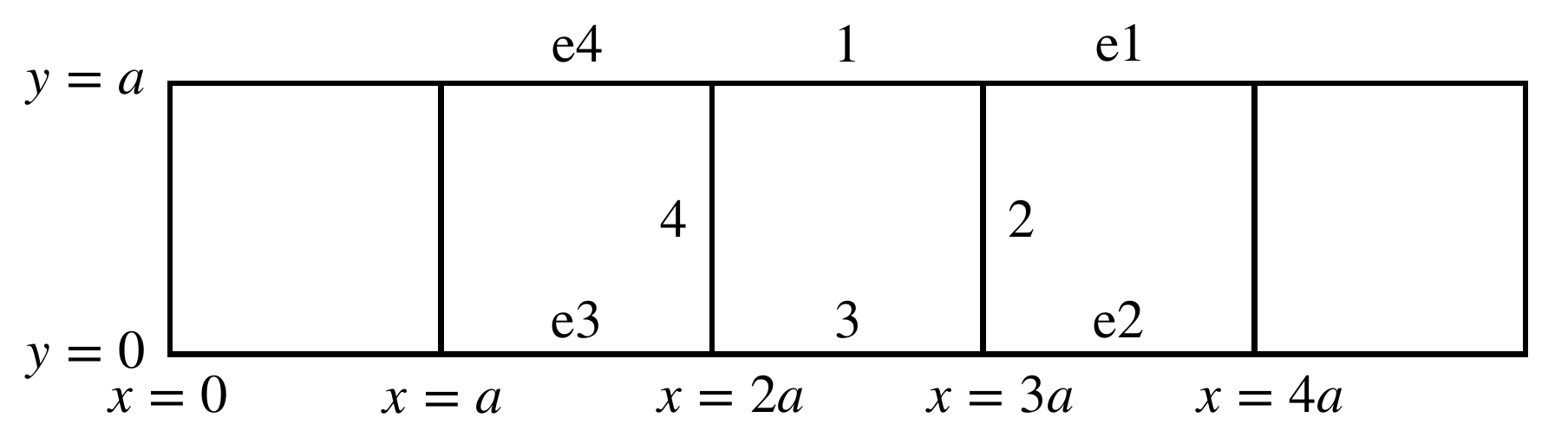}
\caption{A five-plaquette chain. Positions of vertices can be specified by $(x,y)$ as shown. We use the bottom-left corner of a plaquette to specify its position.}
\label{fig:chain}
\end{figure}

We first consider the 2+1D SU(2) pure gauge theory on a plaquette chain and will generalize the discussions to full two and three spatial dimensions in Sec.~\ref{sec:2D3D}. An example plaquette chain is shown in Fig.~\ref{fig:chain}. The bottom-left vertex of each square plaquette is located at $(x=na, y=0)$ where $a$ is the lattice spacing. The Kogut-Susskind Hamiltonian of the theory on a $N$-plaquette chain can be written as~\cite{PhysRevD.11.395}
\begin{align}
\label{eqn:H}
H = \frac{g^2}{2}\sum_{\rm links} E^aE^a - \frac{2}{a^2g^2} \sum_{n=0}^{N-1} \square(n) \,,
\end{align}
where $g$ is the gauge coupling, $E^a$ denotes electric fields on links with $a$ in the superscript standing for SU(2) indexes that are implicitly summed over, and the plaquette operator $\square(n)$ at position $n$ as shown in Fig.~\ref{fig:chain} is defined as
\begin{align}
\square(n) &= \Tr[U^\dagger_y(n,0) U^\dagger_x(n,1) U_y(n+1,0) U_x(n,0)] \,,
\end{align}
where $U_i(n,m)$ is a link variable (Wilson line of length $a$) starting at position $(x=na,y=ma)$ and pointing along the $i$-spatial direction. Physical states in the theory are SU(2) singlets that obey Gauss's law at each vertex $V$
\begin{align}
\label{eqn:gauss}
    \sum_{\ell\in E(V)} E_{\ell}^a = 0\,,\ \forall \, a\ {\rm and}\ V \,,
\end{align}
where $E(V)$ denotes all edges connecting to the vertex $V$.

We represent the lattice Hamiltonian~\eqref{eqn:H} in the electric basis, where the local state on each link is labeled by a quantum number\footnote{We integrate out $m$ quantum numbers on each link~\cite{Klco:2019evd}.} $j$ taking values of integers or half-integers, i.e., $\ket{j}$. The total Hilbert space consists of tensor products of all local link states, i.e, $\otimes_{\ell} \ket{j}_{\ell}$. The electric energy is diagonal in this basis, i.e.,~\cite{Byrnes:2005qx}
\begin{align}
    E^aE^a \ket{j} = j(j+1)  \ket{j}\,,
\end{align}
in which both the electric field and the state are on the same link, otherwise the right-hand side vanishes. The magnetic energy is given by the plaquette operator $\square(n)$ and its matrix element is given by~\cite{Klco:2019evd,ARahman:2021ktn}
\begin{align}
& \bra{J_1J_2J_3J_4}  \square \ket{ j_1j_2j_3j_4 } = \prod_{\alpha=1}^4 \bigg[ (-1)^{j_\alpha+J_\alpha+j_{{\rm e}\alpha}} \nn\\
& \times \sqrt{(2j_\alpha+1)(2J_\alpha+1)}
\left\{ \begin{array}{ccc}  j_{{\rm e}\alpha} & j_\alpha & j_{\alpha\oplus_4 1} \\ 1/2 & J_{\alpha\oplus_4 1} & J_\alpha   \end{array}  \right\}
\bigg] \,,
\end{align}
where the curly bracket denotes the Wigner's 6-j symbol and $\alpha \oplus_4 1 \equiv \alpha+1 \!\! \pmod 4$. $\alpha$ and ${\rm e}\alpha$ refer to the internal and external links of the plaquette under consideration, respectively. An example is shown in Fig.~\ref{fig:chain} for the plaquette whose bottom-left vertex is at $(x=2a,y=0)$.

In numerical calculations, a truncation on the values of the electric flux $j$ is imposed and the maximum value is labeled as $j_{\rm max}$. The generic dependence of $j_{\rm max}$ on accuracy, energy, lattice size and coupling can be found in Appendix A of Ref.~\cite{Turro:2024pxu}, which has also been studied numerically~\cite{Ebner:2023ixq}.
In the following, we will focus on the simplest truncation $j_{\rm max}=1/2$ and construct error correction codes for the SU(2) lattice gauge theory. We assume readers have some familiarity with the stabilizer formalism~\cite{Gottesman:1997zz}.

\section{Code {\romannumeral 1} for $j_{\rm max}=1/2$}
\label{sec:code1}
\subsection{Stabilizers from Gauss's Law}
\begin{figure}[t]
\centering
\includegraphics[width=0.45\textwidth]{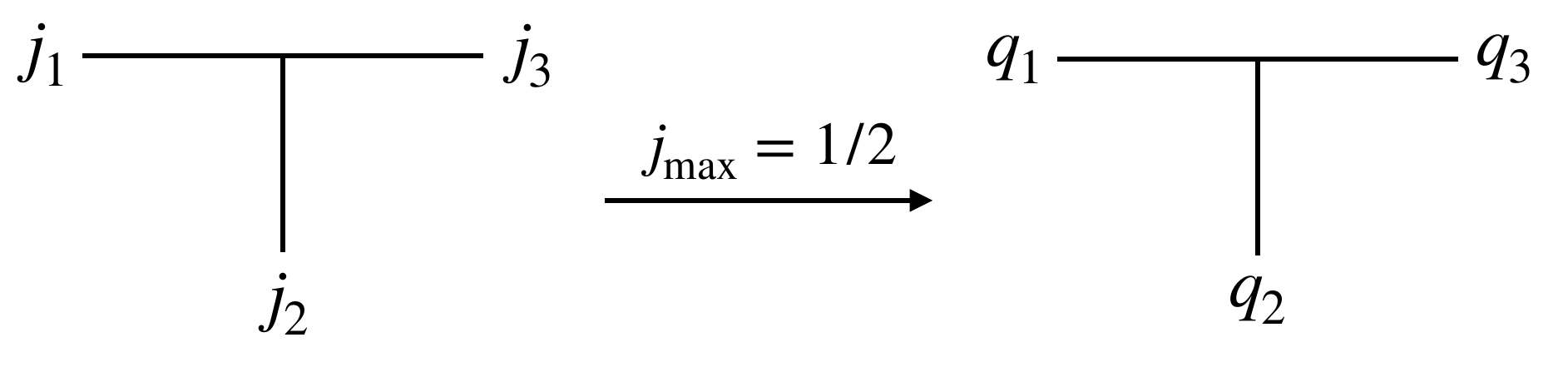}
\caption{A vertex with three links joint. States in the full SU(2) lattice gauge theory are represented by the three $j$ quantum numbers $\ket{j_1j_2j_3}$. When the $j$ quantum number is truncated at $j_{\rm max}=1/2$, the local state on each link becomes a two-level system and can be represented by one qubit. Then the states associated with the vertex can be represented as $\ket{q_1q_2q_3}$. Physical states at $j_{\rm max}=1/2$ are listed in Table~\ref{tab:1} for both ways of representing the states.}
\label{fig:vertex}
\end{figure}

\begin{table}[t]
    \centering
    \begin{tabular}{|c|c|} \hline
       $\ket{j_1j_2j_3}\ $  & \ $\ket{q_1q_2q_3}$  \\ \hline
       $\ket{000}$  & $\ket{000}$ \\ 
       $\ket{\frac{1}{2}\frac{1}{2}0}$  & $\ket{110}$ \\
       $\ket{\frac{1}{2}0\frac{1}{2}}$  & $\ket{101}$ \\
       $\ket{0\frac{1}{2}\frac{1}{2}}$  & $\ket{011}$ \\[0.1cm] \hline
    \end{tabular}
    \caption{Physical states at a vertex when $j_{\rm max}=1/2$ in both representing methods described in Fig.~\ref{fig:vertex}. All the physical states $\ket{q_1q_2q_3}$ are stabilized by $Z_1Z_2Z_3$.}
    \label{tab:1}
\end{table}
We first consider aperiodic boundary conditions, which means the leftmost vertical link is not identified with the rightmost one in Fig.~\ref{fig:chain}. At each vertex, states can be labeled by the $j$ quantum numbers on the three links, i.e., $\ket{j_1j_2j_3}$, as shown in Fig.~\ref{fig:vertex}. When the electric flux $j$ is truncated at $j_{\rm max}=1/2$, each link has only two local states: $j=0$ and $j=1/2$ and thus can be represented by a qubit. States at each vertex then can be represented by three qubits as $\ket{q_1q_2q_3}$. Not all of these states are physical, i.e., obey Gauss's law~\eqref{eqn:gauss}. Only four out of the eight possible states are physical, which are listed in Table~\ref{tab:1}. We find that the physical states are stabilized by the Pauli operator $Z_1Z_2Z_3$, i.e., the physical states have eigenvalue $1$. For the four vertices on the boundary as shown in Fig.~\ref{fig:chain}, states are described by two qubits $\ket{q_1q_2}$ and physical states are stabilized by $Z_1Z_2$.

The collection of the $Z$-stabilizers on all the vertices defines a classical error correction code, which means that it is able to detect and correct any single-qubit $X$-error but not a $Z$-error. For example, if a $X$-error occurs on link 1 in Fig.~\ref{fig:chain}, the two $Z$-stabilizer checks on the left and right ends of link 1 will return $-1$, and vice versa. A $N$-plaquette chain with aperiodic boundary conditions has $3N+1$ links and $2N+2$ vertices. So the system corresponds to $3N+1$ qubits and has $2N+2$ stabilizers, out of which $2N+1$ are independent, since the product of all $Z$-stabilizers at vertices is equal to the identity operator, which physically means that the state is a global SU(2) singlet state. The system encodes $N$ logical classical bits and its distance (the minimal weight of logical operators) is $4$, given by the plaquette operator flipping all the states on its four links.
It follows that this classical error correction code on the $N$-plaquette chain is $[3N+1,N,4]$.

\subsection{Repetition for Quantum Code}
\begin{figure}[t]
\centering
\includegraphics[width=0.45\textwidth]{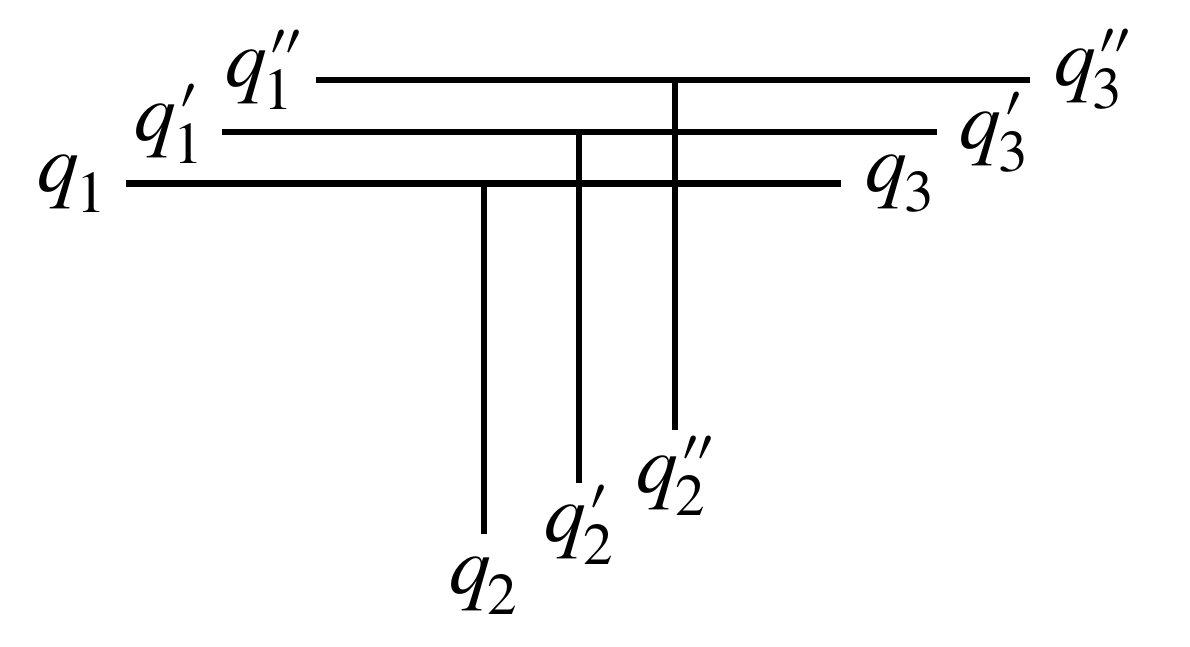}
\caption{A vertex repetition code designed to correct a phase-flip ($Z$) error on any of the nine physical qubits, but only detect a bit-flip ($X$) error, i.e., it is not able to identify the qubit where the $X$-error occurs. The stabilizers are $X_1X_{1'}$, $X_{1'}X_{1''}$, $X_2X_{2'}$, $X_{2'}X_{2''}$, $X_3X_{3'}$, $X_{3'}X_{3''}$ and $Z_1Z_{1'}Z_{1''}Z_2Z_{2'}Z_{2''}Z_3Z_{3'}Z_{3''}$, the last of which corresponds to the local Gauss's law constraint. The syndromes of the $X$-stabilizers for single-qubit $Z$-errors are listed in Table~\ref{tab:2}.
}
\label{fig:qec_at_vertex}
\end{figure}

To make the code quantum, we use the repetition method as done in the Shor's nine-qubit code~\cite{Shor:1995hbe} and the truncated U(1) lattice gauge theory~\cite{Spagnoli:2024mib}. The setup for a vertex is shown in Fig.~\ref{fig:qec_at_vertex}. The local state on each link is repeated three times so the total states associated with the vertex are labeled by nine qubits $\ket{q_1q_{1'}q_{1''}q_2q_{2'}q_{2''}q_3q_{3'}q_{3''}}$. In order to detect and correct any single-qubit $Z$-error, we introduce new stabilizers $X_1X_{1'}$, $X_{1'}X_{1''}$, $X_2X_{2'}$, $X_{2'}X_{2''}$, $X_3X_{3'}$ and $X_{3'}X_{3''}$. 
The original $Z$-stabilizer that originates from Gauss's law now becomes $Z_1Z_{1'}Z_{1''}Z_2Z_{2'}Z_{2''}Z_3Z_{3'}Z_{3''}$. Each vertex involves nine qubits and seven stabilizers. The local codeword (the space of logical states) at one vertex is
\begin{align}
\label{eqn:codeword}
    \mathcal{C}_S &= \{ \ket{\alpha}_1 \ket{\alpha}_2 \ket{\alpha}_3\,,\ \ket{\alpha}_1 \ket{\beta}_2 \ket{\beta}_3\,,\ \ket{\beta}_1 \ket{\alpha}_2 \ket{\beta}_3\,, \nn\\
    &\quad\ \ \ket{\beta}_1 \ket{\beta}_2 \ket{\alpha}_3 \} \,,
\end{align}
where $\ket{\alpha}_i$ and $\ket{\beta}_i$ are states for $\ket{q_iq_{i'}q_{i''}}$ with $i\in\{1,2,3\}$ defined as ($X\ket{\pm} = \pm \ket{\pm}$)
\begin{align}
    \ket{\alpha}_i &\equiv \frac{1}{\sqrt{2}} \big(\ket{+++} + \ket{---} \big) \,,\nn\\
    \ket{\beta}_i &\equiv \frac{1}{\sqrt{2}} \big(\ket{+++} - \ket{---} \big) \,.
\end{align}
We see that the four states in the codeword correspond to the four original physical states listed in Table~\ref{tab:1}.

\begin{table}
    \centering
    \begin{tabular}{|c|c|c|c|c|c|c|} \hline
    \multicolumn{6}{|c|}{Syndromes} & \multirow{2}{*}{Error} \\ \cline{1-6}
       $X_1X_{1'}$ & $X_{1'}X_{1''}$ & $X_2X_{2'}$ & $X_{2'}X_{2''}$ & $X_3X_{3'}$ & $X_{3'}X_{3''}$ &  \\ \hline
    1 & 1 & 1 & 1 & 1 & 1 & No \\ \hline
    $-1$ & 1 & 1 & 1 & 1 & 1 & $Z_1$ \\ \hline
    $-1$ & $-1$ & 1 & 1 & 1 & 1 & $Z_{1'}$ \\ \hline
    1 & $-1$ & 1 & 1 & 1 & 1 & $Z_{1''}$ \\ \hline
    1 & 1 & $-1$ & 1 & 1 & 1 & $Z_2$ \\ \hline
    1 & 1 & $-1$ & $-1$ & 1 & 1 & $Z_{2'}$ \\ \hline
    1 & 1 & 1 & $-1$ & 1 & 1 & $Z_{2''}$ \\ \hline
    1 & 1 & 1 & 1 & $-1$ & 1 & $Z_3$ \\ \hline
    1 & 1 & 1 & 1 & $-1$ & $-1$ & $Z_{3'}$ \\ \hline
    1 & 1 & 1 & 1 & 1 & $-1$ & $Z_{3''}$ \\ \hline
    \end{tabular}
    \caption{Syndromes of single-qubit $Z$-errors in the repetition vertex code. To correct a $Z$-error, one just applies a Pauli-$Z$ gate to the error qubit.}
    \label{tab:2}
\end{table}

\begin{figure}[t]
\centering
\includegraphics[width=0.45\textwidth]{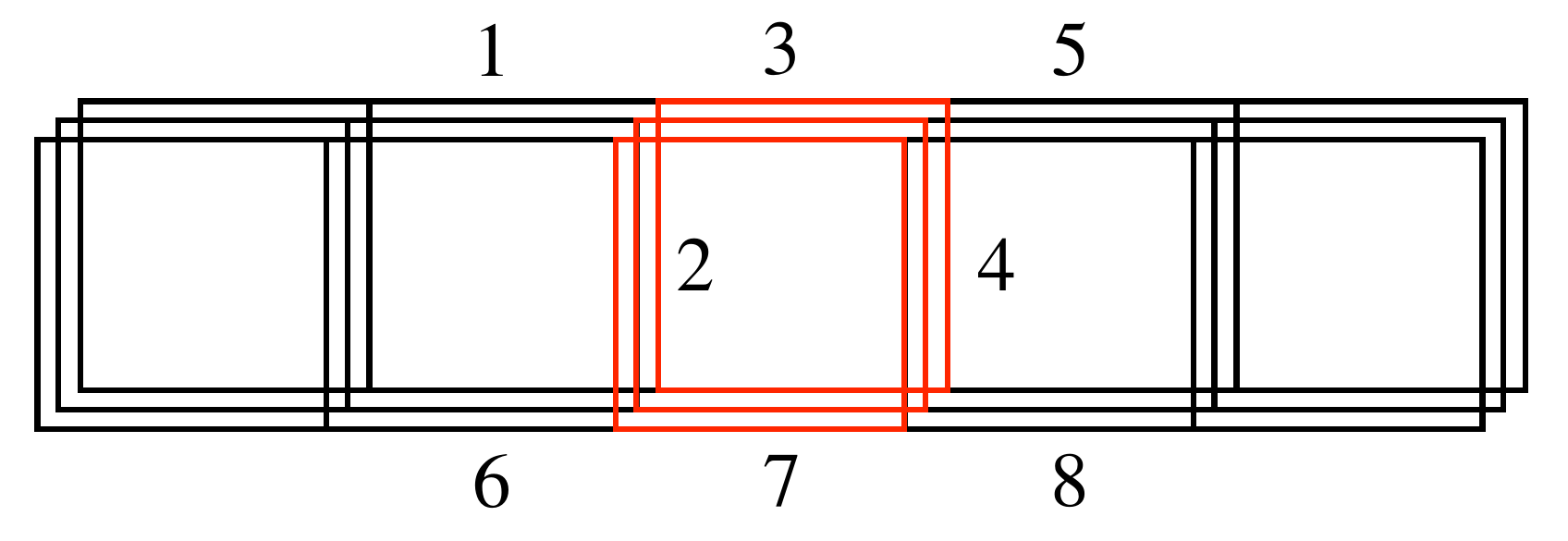}
\caption{Quantum error correction code I for the SU(2) lattice gauge theory on a plaquette chain with aperiodic boundary conditions and $j_{\rm max}=1/2$. Each link is repeated thrice. As mentioned in the caption of Fig.~\ref{fig:qec_at_vertex}, the code is able to correct any single-qubit $Z$-error. Using the $Z^{\otimes 9}$ stabilizers on neighboring vertices, one is also able to detect on which link a single-qubit $X$-error has happened, and thus able to correct it. It is worth noting that the code can identify the $X$-error link but cannot identify which of the three qubits has the error, but is still able to correct it. In other words, the code is degenerate. 
}
\label{fig:qec1_chain}
\end{figure}

A $Z$-error swaps $\ket{+}$ and $\ket{-}$, which can be detected by the $X$-parity checks. 
The syndromes of single-qubit $Z$-errors are listed in Table~\ref{tab:2}. To correct a $Z$-error, one just needs to apply a Pauli-$Z$ gate to the error qubit. 
If an $X$-error occurs on one of the nine qubits, the $Z^{\otimes9}$ stabilizer check will return $-1$. With only this information, we cannot determine where the $X$-error has occurred. With the complete error correction setup for the whole lattice shown in Fig.~\ref{fig:qec1_chain}, we can determine on which link an $X$-error has occurred. For example, if an $X$-error happens on one of the three qubits representing link 2 in Fig.~\ref{fig:qec1_chain}, both the $Z^{\otimes9}$ stabilizer checks at vertex $123$ and vertex $267$ will return $-1$. We still do not know on which one of three qubits for link 2 the $X$-error has happened, but this is already enough for us to correct the error. An $X$-error on a link, no matter on which qubit representing the link, swaps the $\ket{\alpha}$ and $\ket{\beta}$ states. To correct it, we just need to apply a Pauli-$X$ gate to any one of the three qubits. This is why the code is degenerate, just like the Shor's nine-qubit code.

The complete error correction code for the SU(2) lattice gauge theory on a plaquette chain is shown in Fig.~\ref{fig:qec1_chain}. For a $N$-plaquette chain with aperiodic boundary conditions and $j_{\rm max}=1/2$, we have $3N+1$ links on each of which there is a two-level system. With the three-time repetition, the total number of physical qubits is $9N+3$. The number of $X$-stabilizers is $6N+2$ while that of $Z$-stabilizers is $2N+1$ ($2N+2$ vertices but only $2N+1$ stabilizers are independent). So the quantum correction code has $N$ logical qubits and is labeled as $[[9N+3, N, 3]]$. Due to the repetition part of the code, the distance decreases to $3$ compared with the previous classical code. 

\subsection{Logical Gates and Hamiltonian}
The logical $Z$ and $X$ gates for this $[[9N+3, N, 3]]$ code can be chosen to be
\begin{align}
\label{eqn:ZL}
    \overline{Z}(n) &= Z_{t}(n) Z_{t'}(n) Z_{t''}(n) \,, \\
\label{eqn:XL}
    \overline{X}(n) &= X_{t}(n) X_{r}(n) X_{b}(n) X_{l}(n) \,,
\end{align}
where $n$ denotes the $n$th plaquette on the chain and the subscripts $t$, $r$, $b$ and $l$ indicate the top, right, bottom and left links of the $n$th plaquette. For example, the logical $Z$ and $X$ gates for the red plaquette in Fig.~\ref{fig:qec1_chain} are $Z_3Z_{3'}Z_{3''}$ and $X_2X_3X_4X_7$, respectively. We find the logical $Z$ and $X$ gates at the same site anticommute because $Z_{t}(n)$ and $X_{t}(n)$ anticommute. Logical gates at different sites commute. 

It can be seen that logical states have a one-to-one correspondence with plaquette excitations.
In terms of the logical gates, the original SU(2) Hamiltonian on a $N$-plaquette chain with $j_{\rm max}=1/2$ can be written as
\begin{align}
\label{eqn:H_logical}
H &= \frac{3g^2}{2} \sum_{n=0}^{N-1} \frac{1-\overline{Z}(n)}{2} \nn \\
&\quad - \frac{3}{4}g^2\sum_{n=0}^{N-1}\frac{1-\overline{Z}(n)}{2}\frac{1-\overline{Z}(n+1)}{2} \nn \\
&\quad - \frac{2}{a^2g^2} \sum_{n=0}^{N-1} \frac{1+3\overline{Z}(n-1)}{4} \frac{1+3\overline{Z}(n+1)}{4} \overline{X}(n) \,, 
\end{align}
where the aperiodic boundary conditions are imposed by $\overline{Z}(-1)\to 1$ and $\overline{Z}(N)\to 1$. Equation~\eqref{eqn:H_logical} is exactly the spin Hamiltonian found for the physical sector of the SU(2) theory on plaquette chains with $j_{\rm max}=1/2$~\cite{Hayata:2021kcp,ARahman:2022tkr,Yao:2023pht,Illa:2025njz}.

\subsection{Encoding Circuit}
\begin{figure}[h]
$$\Qcircuit @C=2em @R=1.5em {
\lstick{\ket{q_i}=\ket{\psi}} & \gate{H} & \ctrl{1} & \ctrl{2} & \gate{H} & \qw   \\
\lstick{\ket{q_{i'}}=\ket{0}}  & \qw  & \targ & \qw  & \gate{H} & \qw \\
\lstick{\ket{q_{i''}}=\ket{0}}  & \qw  & \qw & \targ  & \gate{H} & \qw} $$
\caption{Encoding circuit in code I for the qubit $q_i$ on a link, where the state can be arbitrary $\ket{\psi}=c_0\ket{0}+c_1\ket{1}$. $H$ in the box stands for the Hadamard gate and the black solid point indicates the control qubit for the CNOT gate. The output state is $c_0\ket{\alpha}_i+c_1\ket{\beta}_i$.}
\label{fig:encode}
\end{figure}
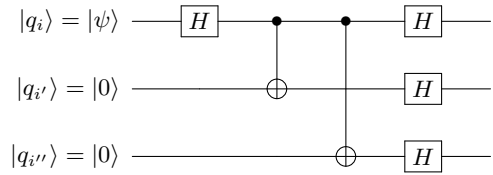

Lastly we discuss how to encode a physical state in the original theory into the quantum correction code. Comparing the physical states at each vertex in the original theory with $j_{\rm max}=1/2$ as listed in Table~\ref{tab:1} and the local codeword in Eq.~\eqref{eqn:codeword}, we see that we just need to encode the $\ket{0}$ state on links in the original theory into the $\ket{\alpha}$ state, and $\ket{1}$ into $\ket{\beta}$. A quantum circuit realizing this for a link is shown in Fig.~\ref{fig:encode}. For the whole lattice, one just needs to apply this encoding circuit to each link. 

\section{Constructions for 2D and 3D}
\label{sec:2D3D}
The above quantum correction code constructed for the SU(2) plaquette chain with $j_{\rm max}=1/2$ can be directly applied to the full 2D plane by using the honeycomb lattice and to 3D space by using either the triamond or the hyperhoneycomb lattice, where each vertex has at most three links attached\footnote{On 2D square and 3D cubic lattices, each vertex has four and six links attached, respectively. SU(2) singlets cannot be uniquely determined by the $j$ quantum numbers on these links as done in Table~\ref{tab:1}, even for $j_{\rm max}=1/2$.}.

\begin{figure}[t]
\centering
\includegraphics[width=0.45\textwidth]{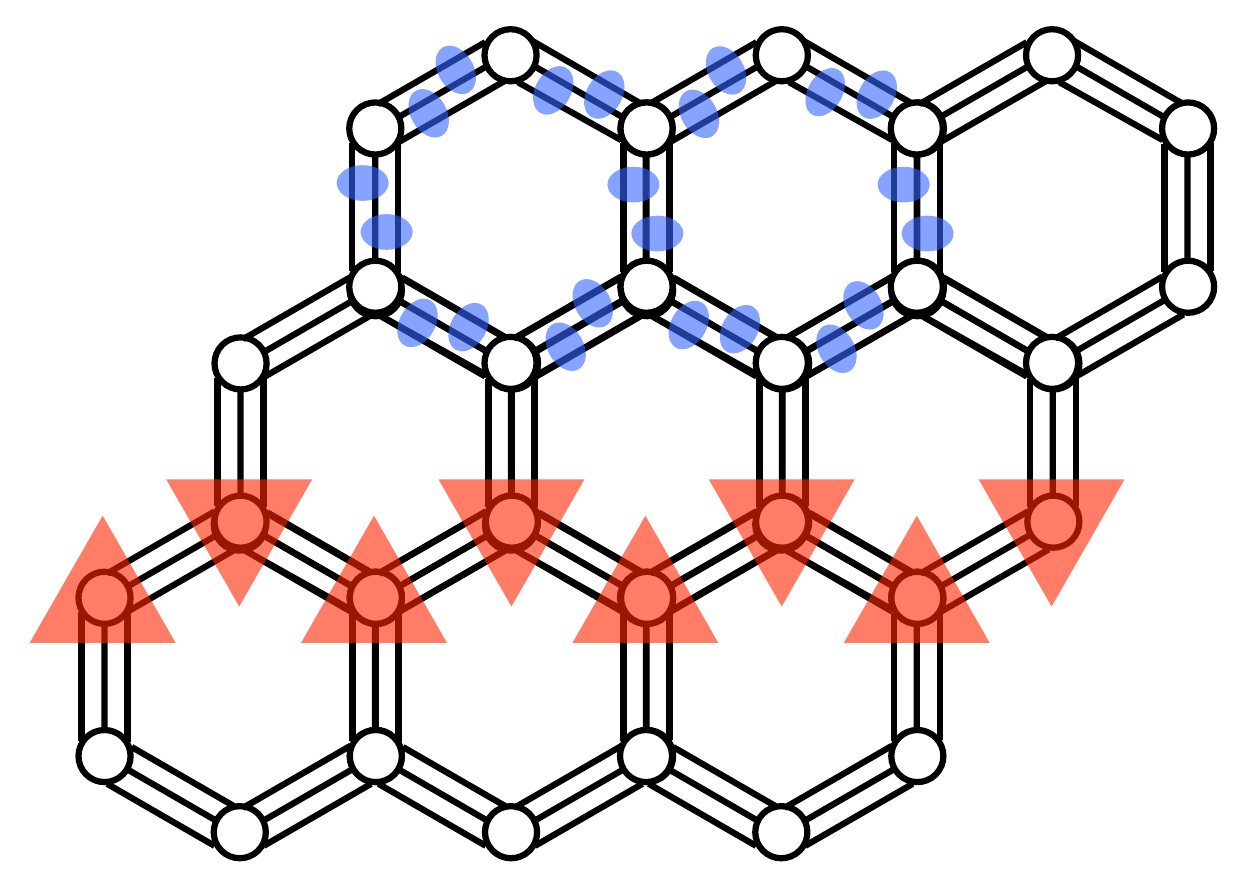}
\caption{Quantum error correction code I for the SU(2) pure gauge theory on a honeycomb lattice with aperiodic boundary conditions and $j_{\rm max}=1/2$, constructed from the local Gauss's law constraint at each vertex and the repetition code. The blue ovals indicate stabilizers of the form $X\otimes X$ on two neighboring repetition links while the red triangles represent the $Z^{\otimes 9}$ or $Z^{\otimes 6}$ stabilizers at each vertex. The code is $[[3(3N_xN_y+2N_x+2N_y-1),N_xN_y,3]]$, where $N_y$ is the number of rows of plaquettes and $N_x$ is the number of plaquettes in each row. In this figure, $N_x=N_y=3$. 
}
\label{fig:qec1_2D}
\end{figure}

The analogous quantum correction code for the honeycomb lattice is shown in Fig.~\ref{fig:qec1_2D}, where each link is repeated three times. The $X$-stabilizers are of the form $X\otimes X$ for neighboring qubits representing each link, some of which are marked as blue ovals in the figure. The $Z$-stabilizers are of the form $Z^{\otimes 9}$ at each vertex ($Z^{\otimes 6}$ on the boundary), some of which are marked as red triangles. For a honeycomb lattice consisting of $N_y$ rows, each of which has $N_x$ plaquettes, the number of vertices is $2(2N_x+1)+(N_y-1)(2N_x+2) = 2N_xN_y+2N_x+2N_y$ and the number of links is $2(2N_x)+N_y(N_x+1)+(N_y-1)(2N_x+1) = 3N_xN_y+2N_x+2N_y-1$. So the total number of qubits used is $3(3N_xN_y+2N_x+2N_y-1)$. The number of $X$-stabilizers is $2(3N_xN_y+2N_x+2N_y-1)$ while the number of $Z$-stabilizers is $2N_xN_y+2N_x+2N_y-1$, since the product of all vertex $Z$-stabilizers is the identity operator and thus one of them is not independent. Then the number of logical qubits is $N_xN_y$ and the quantum error correction code is $[[3(3N_xN_y+2N_x+2N_y-1),N_xN_y,3]]$.

Logical $Z$ and $X$ gates can be analogously written out as Eqs.~\eqref{eqn:ZL} and~\eqref{eqn:XL}, respectively, i.e., the $\overline{Z}$ gate is the product of the three Pauli-$Z$ gates acting on the three qubits of the same link, respectively, while the $\overline{X}$ gate is the product of the six Pauli-$X$ gates acting on the six qubits of the same repetition type around a honeycomb plaquette. The Hamiltonian of the SU(2) gauge theory on the honeycomb lattice with aperiodic boundary conditions and $j_{\rm max}=1/2$ can be written out explicitly in terms of these logical gates and the resulting Hamiltonian is exactly the spin Hamiltonian found in Refs.~\cite{Muller:2023nnk,Illa:2025njz}.

Using the same method, we can construct a similar quantum correction code for the SU(2) theory with $j_{\rm max}=1/2$ on the 3D triamond and the hyperhoneycomb lattices. We omit the technical details here.

\section{Periodic Boundary Conditions and Topological Flux Configurations}
\label{sec:pbc}
So far we have been focusing on lattices with aperiodic boundary conditions. Now we discuss the case of periodic boundary conditions. 

\begin{figure}[b]
\centering
\includegraphics[width=0.45\textwidth]{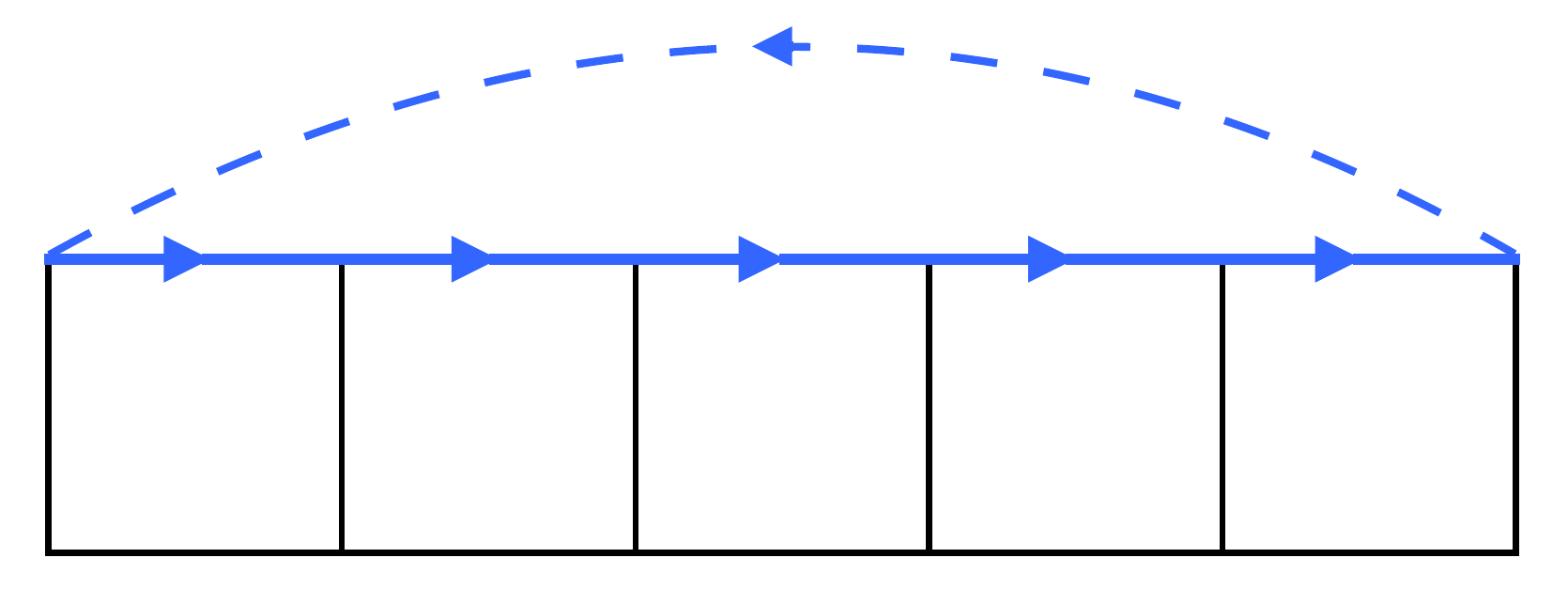}
\caption{Electric flux winding around a five-plaquette chain with periodic boundary conditions. The links in blue are in the state $\ket{j=1/2}$ (equivalently $\ket{q}=\ket{0}$) while those in black are in the state $\ket{j=0}$ ($\ket{q}=\ket{1}$).}
\label{fig:flux}
\end{figure}

On a $N$-plaquette chain with periodic boundary conditions and $j_{\rm max}=1/2$, the number of links is $3N$ and the number of vertices is $2N$. After the repetition, the number of physical qubits is $9N$. The number of $X$-stabilizers is $6N$ while the number of independent $Z$-stabilizers originating from Gauss's law is $2N-1$. So the error correction code is $[[9N,N+1,3]]$. For the same number of plaquettes, the lattice with periodic boundary conditions has one more logical qubit degrees of freedom than the lattice with aperiodic boundary conditions. This extra qubit degree of freedom describes whether the system has one unit of topological electric flux winding around the chain. Due to the $j_{\rm max}=1/2$ truncation, this topological flux can only be one unit or none. An example of states with such a topological electric flux is shown in Fig.~\ref{fig:flux}, which cannot be obtained from the bare vacuum state that has $\ket{j=0}$ on all links by applying local plaquette terms in the Hamiltonian.
On the other hand, the state with all the top and bottom links in the $\ket{j=1/2}$ states and everywhere else in the $\ket{j=0}$ states can be obtained from the bare vacuum state by applying local plaquette operators, and thus is a topologically trivial state. The logical $X$ gate corresponding to this topological excitation is given by, e.g., the tensor product of all Pauli-$X$ gates acting on the qubits of the same repetition type in the bottom links 
\begin{align}
    \overline{X}^{({\rm top})} = \bigotimes_{n=0}^{N-1} X_b(n) \,,
\end{align}
which can be seen to commute with all the stabilizers and Eqs.~\eqref{eqn:ZL} and~\eqref{eqn:XL}. We also note that $\overline{X}^{({\rm top})}$ does not commute with the $Z^{\otimes 6}$ stabilizers at the bottom-left and bottom-right corners of an aperiodic chain. In other words, the topological flux configurations do not exist on aperiodic chains.

On a $N_x\times N_y$ periodic honeycomb lattice with $j_{\rm max}=1/2$, we have two types of topological electric fluxes: one along the horizontal direction and the other along the vertical direction. We check if the number of logical qubit degrees of freedom works out. The number of vertices is $2N_xN_y$ and the number of links is $3N_xN_y$. So the quantum error correction code is $[[9N_xN_y, N_xN_y+1,3]]$. Compared with the aperiodic lattice, the periodic one has one extra logical qubit degree of freedom. But we expect two additional qubit degrees of freedom, which correspond to the two types of topological electric fluxes. This is because on a periodic honeycomb lattice, there is an emergent global spin flip symmetry~\cite{Muller:2023nnk}. In a given topological sector, flipping all the logical qubits gives the same state in the original $j$ representation. This symmetry is absent on aperiodic honeycomb lattices. Removing this degeneracy on the periodic lattice reduces the number of qubit degrees of freedom by one. In the end, we only see one net additional qubit degree of freedom.

Topological electric fluxes on periodic triamond and hyperhoneycomb lattices can be similarly studied.

\section{Code {\romannumeral 2} for $j_{\rm max}=1/2$}
\label{sec:code2}
The physical states at a vertex listed in Table~\ref{tab:1} are stabilized by $Z^{\otimes 3}$, but not by any $X$-operator. This is why we have to use the repetition method to make the code robust against single-qubit $Z$-errors. It is possible to make the physical states also stabilized by an $X$-operator by adding an extra qubit to the vertex. The correspondence between the physical states at a vertex in the original theory and the four-qubit representation is given in Table~\ref{tab:3}. The physical states are now stabilized by both $Z_1Z_2Z_3Z_4$ and $X_1X_2X_3X_4$. This code uses four qubits to encode two logical qubits, and  able to detect single-qubit $Z$-error or $X$-error but not able to correct it. Thus, it is the quantum code $[[4,2,2]]$~\cite{Grassl:1996eh,Vaidman:1996bg}. 

\begin{table}[th]
    \centering
    \begin{tabular}{|c|c|} \hline
       $\ket{j_1j_2j_3}\ $  & \ $\ket{q_1q_2q_3q_4}$  \\ \hline
       $\ket{000}$  & $\frac{1}{\sqrt{2}}(\, \ket{0000} + \ket{1111}\,)$ \\ 
       $\ket{\frac{1}{2}\frac{1}{2}0}$  & $\frac{1}{\sqrt{2}}(\, \ket{1100} + \ket{0011}\,)$ \\
       $\ket{\frac{1}{2}0\frac{1}{2}}$  & $\frac{1}{\sqrt{2}}(\, \ket{1010} + \ket{0101}\,)$ \\
       $\ket{0\frac{1}{2}\frac{1}{2}}$  & $\frac{1}{\sqrt{2}}(\, \ket{0110} + \ket{1001}\,)$ \\[0.1cm] \hline
    \end{tabular}
    \caption{Physical states at a vertex when $j_{\rm max}=1/2$ in the original theory and the correspondent four-qubit representations. All the physical states $\ket{q_1q_2q_3q_4}$ are stabilized by both $Z_1Z_2Z_3Z_4$ and $X_1X_2X_3X_4$.}
    \label{tab:3}
\end{table}

To make the $[[4,2,2]]$ code able to correct any single-qubit error, we concatenate it with the $C_6=[[6,2,2]]$ code~\cite{Knill:2004apf}. In particular, we introduce two more copies of the vertex and end up with 12 qubits $\ket{q_1 q_{1'} q_{1''} q_2 q_{2'} q_{2''} q_3 q_{3'} q_{3''} q_4 q_{4'} q_{4''} }$. The stabilizers inherited from the $[[4,2,2]]$ code are
\begin{align}
\label{eqn:stabilizers422}
&Z_1Z_2Z_3Z_4 \,,&   &Z_{1'}Z_{2'}Z_{3'}Z_{4'} \,,& &Z_{1''}Z_{2''}Z_{3''}Z_{4''} \,,& \nn\\
&X_1X_2X_3X_4 \,,&   &X_{1'}X_{2'}X_{3'}X_{4'} \,,& &X_{1''}X_{2''}X_{3''}X_{4''} \,,& 
\end{align}
while those from the $C_6$ code are
\begin{align}
&Z_1Z_{1'}Z_{1''}Z_2Z_{2'}Z_{2''} \,, & &Z_2Z_{2'}Z_{2''}Z_3Z_{3'}Z_{3''} \,, &\nn\\
&X_1X_{1'}X_{2'}X_{2''}X_{3''}X_3 \,, & &X_2X_{2'}X_{3'}X_{3''}X_{1''}X_1 \,. &
\end{align}
The resulting quantum error correction code is the carbon code $[[12,2,4]]$~\cite{Paetznick:2024ztu}, which uses 12 qubits to encode two logical qubits and is able to correct any single-qubit error.

\begin{widetext}
The check matrix for these stabilizers is given by
\begin{align}
H_s = \left[\begin{array}{cccccccccccc|cccccccccccc}
1 & 0 & 0 & 1 & 0 & 0 & 1 & 0 & 0 & 1 & 0 & 0 & 0 & 0 & 0 & 0 & 0 & 0 & 0 & 0 & 0 & 0 & 0 & 0 \\
0 & 1 & 0 & 0 & 1 & 0 & 0 & 1 & 0 & 0 & 1 & 0 & 0 & 0 & 0 & 0 & 0 & 0 & 0 & 0 & 0 & 0 & 0 & 0 \\
0 & 0 & 1 & 0 & 0 & 1 & 0 & 0 & 1 & 0 & 0 & 1 & 0 & 0 & 0 & 0 & 0 & 0 & 0 & 0 & 0 & 0 & 0 & 0 \\
1 & 1 & 0 & 0 & 1 & 1 & 1 & 0 & 1 & 0 & 0 & 0 & 0 & 0 & 0 & 0 & 0 & 0 & 0 & 0 & 0 & 0 & 0 & 0 \\
1 & 0 & 1 & 1 & 1 & 0 & 0 & 1 & 1 & 0 & 0 & 0 & 0 & 0 & 0 & 0 & 0 & 0 & 0 & 0 & 0 & 0 & 0 & 0 \\ \hline
0 & 0 & 0 & 0 & 0 & 0 & 0 & 0 & 0 & 0 & 0 & 0 & 1 & 0 & 0 & 1 & 0 & 0 & 1 & 0 & 0 & 1 & 0 & 0 \\
0 & 0 & 0 & 0 & 0 & 0 & 0 & 0 & 0 & 0 & 0 & 0 & 0 & 1 & 0 & 0 & 1 & 0 & 0 & 1 & 0 & 0 & 1 & 0 \\
0 & 0 & 0 & 0 & 0 & 0 & 0 & 0 & 0 & 0 & 0 & 0 & 0 & 0 & 1 & 0 & 0 & 1 & 0 & 0 & 1 & 0 & 0 & 1 \\
0 & 0 & 0 & 0 & 0 & 0 & 0 & 0 & 0 & 0 & 0 & 0 & 1 & 1 & 1 & 1 & 1 & 1 & 0 & 0 & 0 & 0 & 0 & 0 \\
0 & 0 & 0 & 0 & 0 & 0 & 0 & 0 & 0 & 0 & 0 & 0 & 0 & 0 & 0 & 1 & 1 & 1 & 1 & 1 & 1 & 0 & 0 & 0
\end{array} \right] \,,
\end{align}
where each row in the top left block corresponds to the $X$-stabilizers while that in the bottom right block originates from the $Z$-stabilizers. If the stabilizer contains a $X$ or $Z$ operator acting on a qubit, we put $1$ at the corresponding entry. Otherwise we put $0$. By Gaussian elimination procedure, we can transform the check matrix $H_s$ to a standard form defined by the block structure~\cite{nielsen2010quantum} (here $r=5$, $n=12$ and $k=2$)
\begin{align}
H_s = \left[\begin{array}{ccc|ccc}
I_{r\times r} & A^1_{r\times (n-k-r)} & A^2_{r\times k} & B_{r\times r} & \mathbf{0}_{r\times(n-k-r)} & C_{r\times k} \\ \hline
\mathbf{0}_{(n-k-r)\times r} & \mathbf{0}_{(n-k-r)\times (n-k-r)} & \mathbf{0}_{(n-k-r)\times k} & D_{(n-k-r)\times r} & I_{(n-k-r)\times(n-k-r)} & E_{(n-k-r)\times k} 
\end{array}
\right] \,,
\end{align}
where $I$ and $\mathbf{0}$ indicate identity and zero matrices, respectively and $A^1, A^2, B,C,D,E$ are block matrices. In particular, we obtain
\begin{align}
\label{eqn:checkM}
H_s = \left[\begin{array}{cccccccccccc|cccccccccccc}
1 & 0 & 0 & 0 & 0 & 1 & 1 & 1 & 1 & 0 & 1 & 0 & 0 & 0 & 0 & 0 & 0 & 0 & 0 & 0 & 0 & 0 & 0 & 0 \\
0 & 1 & 0 & 0 & 0 & 1 & 1 & 0 & 0 & 1 & 1 & 1 & 0 & 0 & 0 & 0 & 0 & 0 & 0 & 0 & 0 & 0 & 0 & 0 \\
0 & 0 & 1 & 0 & 0 & 1 & 0 & 0 & 1 & 0 & 0 & 1 & 0 & 0 & 0 & 0 & 0 & 0 & 0 & 0 & 0 & 0 & 0 & 0 \\
0 & 0 & 0 & 1 & 0 & 1 & 0 & 1 & 1 & 1 & 1 & 0 & 0 & 0 & 0 & 0 & 0 & 0 & 0 & 0 & 0 & 0 & 0 & 0 \\
0 & 0 & 0 & 0 & 1 & 1 & 1 & 1 & 0 & 1 & 0 & 1 & 0 & 0 & 0 & 0 & 0 & 0 & 0 & 0 & 0 & 0 & 0 & 0 \\ \hline
0 & 0 & 0 & 0 & 0 & 0 & 0 & 0 & 0 & 0 & 0 & 0 & 1 & 1 & 1 & 1 & 1 & 1 & 0 & 0 & 0 & 0 & 0 & 0 \\
0 & 0 & 0 & 0 & 0 & 0 & 0 & 0 & 0 & 0 & 0 & 0 & 0 & 1 & 1 & 1 & 0 & 0 & 1 & 0 & 0 & 0 & 1 & 1 \\
0 & 0 & 0 & 0 & 0 & 0 & 0 & 0 & 0 & 0 & 0 & 0 & 0 & 1 & 0 & 0 & 1 & 0 & 0 & 1 & 0 & 0 & 1 & 0 \\
0 & 0 & 0 & 0 & 0 & 0 & 0 & 0 & 0 & 0 & 0 & 0 & 1 & 1 & 0 & 1 & 1 & 0 & 0 & 0 & 1 & 0 & 0 & 1 \\
0 & 0 & 0 & 0 & 0 & 0 & 0 & 0 & 0 & 0 & 0 & 0 & 1 & 1 & 1 & 0 & 0 & 0 & 0 & 0 & 0 & 1 & 1 & 1
\end{array} \right] \,.
\end{align}
\end{widetext}

From the check matrix in the standard form, the logical $Z$-gates are given by the rows of~\cite{nielsen2010quantum}
\begin{align}
\left[\begin{array}{ccc|ccc}
\mathbf{0}_{k\times r} & \mathbf{0}_{k\times (n-k-r)} & \mathbf{0}_{k\times k} & (A^2)^T & \mathbf{0}_{k\times (n-k-r)} & I_{k\times k} 
\end{array}
\right] \,.
\end{align}
while the logical $X$-gates are given by the rows of 
\begin{align}
\label{eqn:logical-X-generic}
\left[\begin{array}{ccc|ccc}
\mathbf{0}_{k\times r} & E^T & I_{k\times k} & C^T & \mathbf{0}_{k\times (n-k-r)} & \mathbf{0}_{k\times k} 
\end{array}
\right] \,.
\end{align}
The left half gives $X$-gate contents while the right half gives $Z$-gate components, just as in the check matrix.
One can show these logical gates commute with all the stabilizers and satisfy correct commutation relations among themselves.
For the $[[12,2,4]]$ code, we find
\begin{align}
    &\overline{Z}_1 = Z_1Z_{1'}Z_2Z_{4'} \,, & & \overline{Z}_2 = Z_{1'}Z_{1''}Z_{2'}Z_{4''} \,, &\nn\\
    &\overline{X}_1 = X_3X_{3'}X_4X_{4'} \,, & & \overline{X}_2 = X_3X_{3''}X_4X_{4''} \,. &
\end{align}

\begin{figure*}[t]
\centering
\includegraphics[width=0.9\textwidth]{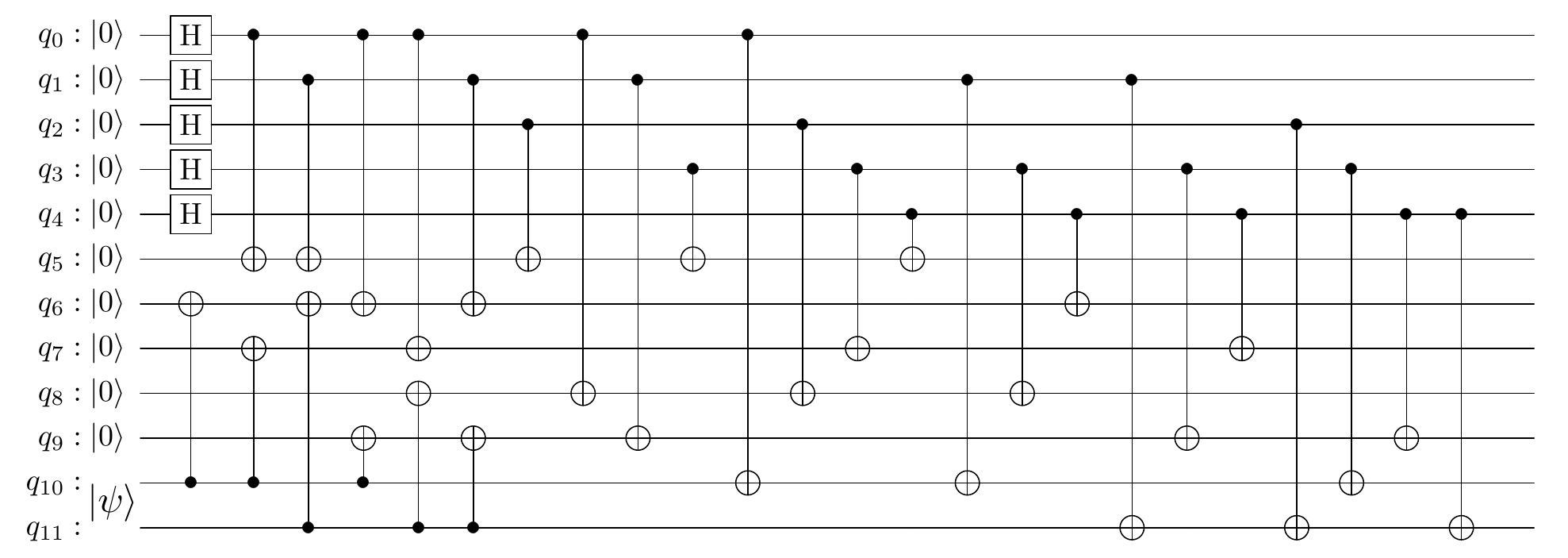}
\caption{Encoding circuit in code II for the four physical states at a vertex. The original state $\ket{\psi}$ in this four dimensional Hilbert space is taken as an input on the last two qubits. All the other qubits are initialized to the $\ket{0}$ state.}
\label{fig:encode2}
\end{figure*}

Using the method pioneered by Gottesman~\cite{Gottesman:1997zz}, we can construct the encoding circuit from the check matrix~\eqref{eqn:checkM} and the logical $X$-operators, which is shown in Fig.~\ref{fig:encode2}. The qubits 0--11 in the circuit are ordered to match the qubit sequence in the sentence just above Eq.~\eqref{eqn:stabilizers422}.
We use the qubit on the left link and the qubit on the right link as in Fig.~\ref{fig:vertex} and Table~\ref{tab:1} to label the four physical states in the original theory, so a generic physical state can be written as $\ket{\psi} = c_{ij}\ket{ij} = c_{00} \ket{000} + c_{01} \ket{0\frac{1}{2}\frac{1}{2}} + c_{10} \ket{\frac{1}{2}\frac{1}{2}0} + c_{11} \ket{\frac{1}{2}0\frac{1}{2}}$. It lives in a four-dimensional Hilbert space and is taken as an input on the last two qubits in the encoding circuit, where the rest of qubits are initialized to the $\ket{0}$ state. We use the same notation for logical states, i.e., $\ket{q_lq_r}_L$ where $q_l$ labels the state on the left link of the vertex while $q_r$ on the right link.
In particular, the logical state $\ket{00}_L$ corresponds to $\ket{000}$ in the original $j$ representation, $\ket{01}_L$ to $\ket{0\frac{1}{2}\frac{1}{2}}$, $\ket{10}_L$ to $\ket{\frac{1}{2}\frac{1}{2}0}$ and $\ket{11}_L$ to $\ket{\frac{1}{2}0\frac{1}{2}}$.

The circuit is constructed from the expression of logical states~\cite{Mondal:2023dgi}
\begin{align}
    |q_lq_r\rangle_L \propto \overline{X}_1^{q_l} \overline{X}_2^{q_r} \prod_{j=1}^{10} \frac{1+g_j}{\sqrt{2}} |000\, 000\, 000\, 000\rangle \,,
\end{align}
where $g_j$ denotes a stabilizer. A $Z$-stabilizer does not change the trivial state $|000\, 000\, 000\, 000\rangle$, so we do not need to consider them. The six CNOT gates in the lower left corner of Fig.~\ref{fig:encode2} that have the input qubits as controls originate from the application of $\overline{X}_1^{q_l} \overline{X}_2^{q_r}$. They do not involve qubits 0--5 because $\overline{X}_1$ and $\overline{X}_2$ do not act on them. The five Hadamard gates on qubits 0--4 and the followed control operations implement the five $X$-stabilizer projections, since the $X$-stabilizers in the standard form of the check matrix have an identity block [see the $I_{r\times r}$ in Eq.~\eqref{eqn:checkM}]. Note in both cases, if a control qubit is already $1$, no $X$-gate is applied on it even if the logical-$X$ operator or the $X$-stabilizer contains such an $X$-gate.

\begin{figure}[t]
\centering
\includegraphics[width=0.48\textwidth]{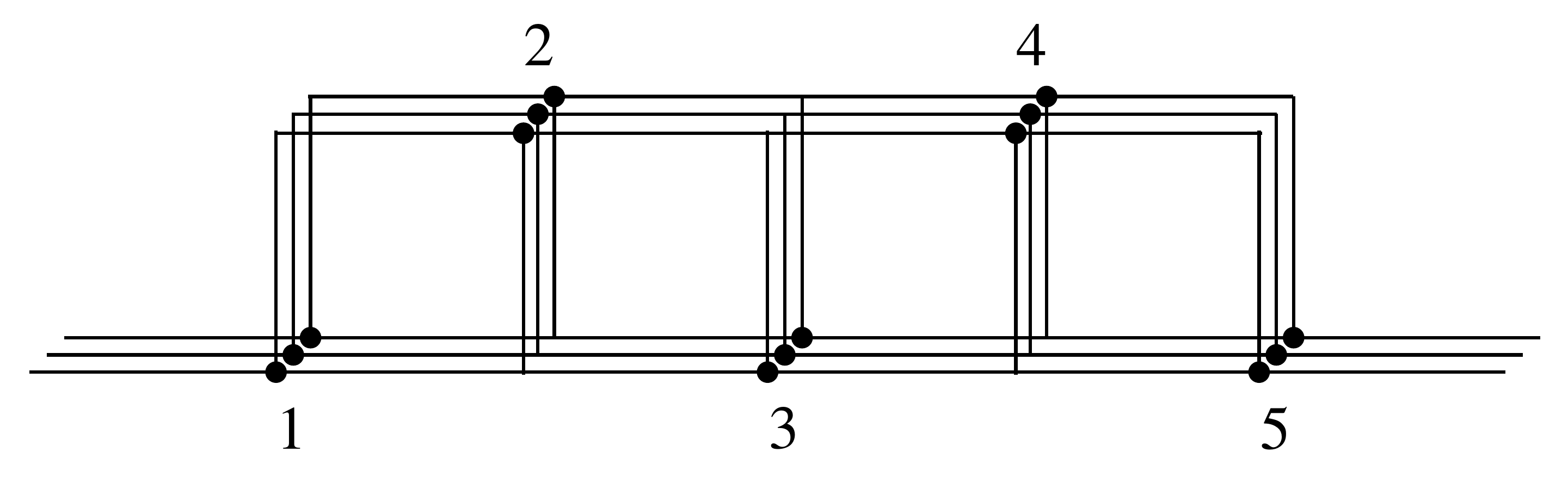}
\caption{Quantum error correction code II for the SU(2) lattice gauge theory on a plaquette chain. The code converts local Gauss's law constraints at half of the vertices (marked as black dots) into stabilizers. The constraints at the other half of the vertices are not imposed. Thus, the Hilbert space of this code contains unphysical states. It also contains the topological flux configuration mentioned in Sec.~\ref{sec:pbc} if the boundary conditions are periodic.}
\label{fig:qec2_chain}
\end{figure}

This error correction code for the whole plaquette chain is shown in Fig.~\ref{fig:qec2_chain}, in which black dots indicate the vertices where the local Gauss's law constraints have been converted into stabilizers while the constraints at vertices without black dots are not accounted for in this code. Thus the logical Hilbert space of this code still contains some unphysical states that violate Gauss's law at vertices without black dots. As long as the initial state is physical and the Hamiltonian evolution is implemented correctly, the state will remain in the physical sector. If the boundary conditions are periodic, the codeword also contains the topological flux configuration mentioned in Sec.~\ref{sec:pbc}.

In terms of the logical operators, the total electric energy is given by
\begin{align}
\label{eqn:HE}
    H_E = \frac{3g^2}{16}\sum_n \big[ 3 - \overline{Z}_1(n) - \overline{Z}_2(n) - \overline{Z}_1(n) \overline{Z}_2(n) \big] \,,
\end{align}
in which $n$ labels the positions of the black-dot vertices in Fig.~\ref{fig:qec2_chain}, where local Gauss's law constraints have been converted into stabilizers. The local magnetic interaction term (the plaquette) involves four black-dot vertices
\begin{align}
\square(n) &\to \overline{M}_2(n-1)\overline{M}_1(n)\overline{M}_2(n+1)\overline{M}_1(n+2) \nn\\
& \quad \otimes \overline{X}_2(n) \overline{X}_1(n+1) \,,
\end{align}
where we have defined
\begin{align}
\label{eqn:M}
    \overline{M}_i(n) = \frac{1+\overline{Z}_i(n)}{2} + \frac{1-\overline{Z}_i(n)}{2\sqrt{2}} \,.
\end{align}

\begin{figure}[t]
\centering
\includegraphics[width=0.48\textwidth]{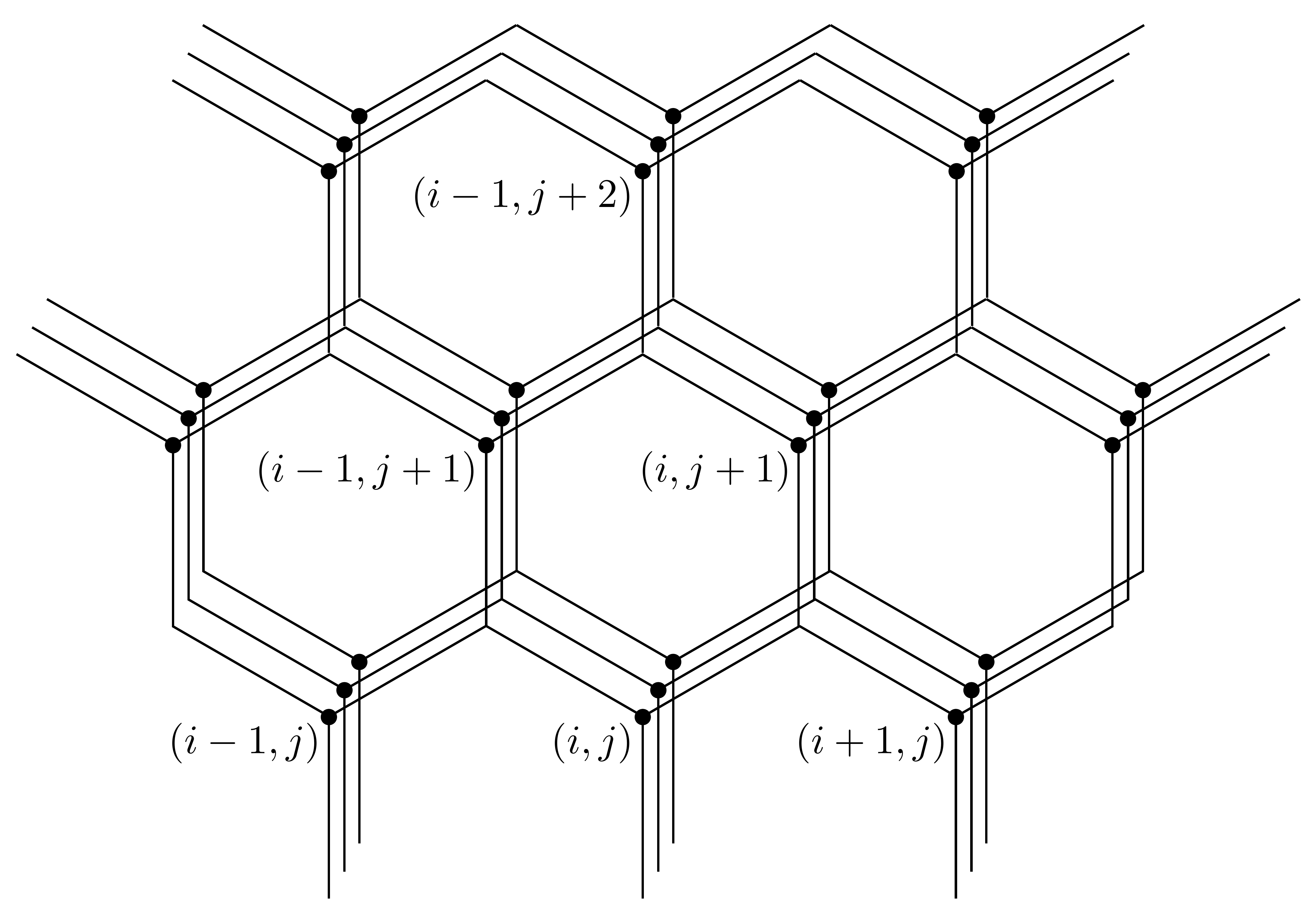}
\caption{Quantum error correction code II for the SU(2) lattice gauge theory on a honeycomb lattice. The vertices where local Gauss's laws have been converted into stabilizers are labeled as black dots. Their positions are indicated by $(i,j)$. }
\label{fig:qec2_2D}
\end{figure}

Extending this code to two spatial dimensions is straightforward on a honeycomb lattice, as shown in Fig.~\ref{fig:qec2_2D}. At each black-dot vertex, we still use the notation that the two logical qubits $\ket{q_lq_r}_L$ represent the $j$ states on the left and right links of the vertex. The local electric energy associated with each black-dot vertex is the same as in Eq.~\eqref{eqn:HE} except for a different prefactor, accounting for the geometry of the unit cell. The local magnetic interaction term (the honeycomb plaquette) involves six black-dot vertices
\begin{align}
    \varhexagon(i,j) &\to \overline{M}_2(i-1,j) \overline{M}_{12}(i,j) \overline{M}_1(i+1,j) \nn\\
    &\quad \otimes 
    \overline{M}_1(i-1,j+1) \overline{M}_2(i,j+1) \overline{M}_{12}(i-1,j+2) \nn\\
    &\quad \otimes \overline{X}_1(i,j) \overline{X}_2(i,j) \overline{X}_2(i-1,j+1) \overline{X}_1(i,j+1) \,,
\end{align}
where we used an analog of Eq.~\eqref{eqn:M} and defined 
\begin{align}
    \overline{M}_{12}(i,j) = \frac{1+\overline{Z}_1(i,j)\overline{Z}_2(i,j)}{2} + \frac{1-\overline{Z}_1(i,j)\overline{Z}_2(i,j)}{2\sqrt{2}} \,.
\end{align}

The same code is directly applicable to the 3D triamond and hyperhoneycomb lattices, since these lattices only have vertices with at most three links attached. The only complication is that the local plaquette term involves more logical operators. We omit the technical details here.

\section{Conclusions}
\label{sec:conclusions}
In this paper, we constructed two quantum error correction codes for pure SU(2) gauge theory on lattices with only vertices to which at most three links are attached, such as square plaquette chains, honeycombs, triamonds and hyperhoneycombs in the electric basis truncated at $j_{\rm max}=1/2$. For a lattice with $N$ plaquettes, code I converts local Gauss's law constraints at all vertices into stabilizers and roughly uses $9N$ physical qubits to encode $N$ logical qubits while code II only converts half constraints and uses $12N$ physical qubits to encode $2N$ logical qubits. The Hilbert space of code II also contains unphysical states violating Gauss's law at vertices that are not used as stabilizers. Both codes can correct any single-qubit error. If an error occurs on one of the links attached to a vertex, code I requires stabilizer check results on nearby vertices in order to identify and correct the error, while code II only needs check results associated with the local vertex. Code II is locally the carbon code.

As a comparison, if one uses the five-qubit perfect code~\cite{Laflamme:1996iw} to encode the $j$ state on each link, which is able to correct any single-qubit error, it will take about $15N$ physical qubits for a $N$-plaquette lattice. This shows converting Gauss's law constraints in the truncated SU(2) lattice gauge theory into stabilizers reduces the qubit cost for quantum error correction, as it utilizes the native redundancy in the theory. However, it is worth pointing out that, if one uses the spin Hamiltonian for physical states in the $j_{\rm max}=1/2$ SU(2) lattice gauge theory as the starting point, in which Gauss's law at each vertex has been ``integrated out'', one will only need a $N$-qubit system to describe the physical Hilbert space. Then using the five-qubit perfect code to conduct error correction will only need $5N$ physical qubits. Despite the efficiency, this approach requires that one is able to solve the Gauss's law constraints at all vertices and write down a Hamiltonian in the physical Hilbert space, which seems very difficult in general. On the other hand, as we showed in the construction of code I, the stabilizer formalism of quantum error correction can guide us to find out such a physical Hamiltonian.

Constructing an error correction code for $j_{\rm max}>1/2$ will be useful for future quantum simulations of pure SU(2) lattice gauge theory, in particular for taking the continuum limit. For a vertex with three links in the state $j_1$, $j_2$, and $j_3$, respectively, the local Gauss's law constraint can be mathematically formulated as
\begin{align}
\label{eqn:gauss_general}
    |j_1-j_2| \leq j_3 \leq j_1+j_2 \,, \quad j_1+j_2+j_3 \in \mathbb{Z}_{\geq 0} \,.
\end{align}
How to convert Eq.~\eqref{eqn:gauss_general} into one or more stabilizers will be explored in future work. Finally, it will be worth understanding how to systematically enlarge the distance in these error correction codes such that they can correct multi-qubit errors, which are crucial for fault-tolerant quantum simulations.

\begin{acknowledgments}
We would like to thank Alessandro Roggero for useful discussions. 
We acknowledge the workshop ``Co-design for Fundamental Physics in the Fault-Tolerant Era" held at the InQubator for Quantum Simulation (IQuS)\footnote{\url{https://iqus.uw.edu/}} hosted by the Institute for Nuclear Theory in April 2025.
This work is supported by the U.S. Department of Energy, Office of Science, Office of Nuclear Physics, IQuS under Award Number DOE (NP) Award No. DE-SC0020970 via the program on Quantum Horizons: QIS Research and Innovation for Nuclear Science.
\end{acknowledgments}

\bibliography{main.bib}
\end{document}